\documentclass[a4paper,11pt]{article}
\pdfoutput=1 

\usepackage{jcappub} 
\usepackage[T1]{fontenc} 
\usepackage[utf8]{inputenc}
\usepackage{soul}

\newcommand{\sv}{\langle \sigma v \rangle}
\newcommand{\Jf}{$\mathcal{J}$-factor}

\newcommand{\refmorf}[1]{{\it BjX}}

\title{\boldmath Particle Dark Matter Constraints: the Effect of Galactic Uncertainties}

\author[a]{Maria Benito,}
\author[a,b]{Nicol\'as Bernal,}

\author[c]{Nassim Bozorgnia,}
\author[c,d]{Francesca Calore}
\author[a]{and Fabio Iocco}

\affiliation[a]{ICTP South American Institute for Fundamental Research\\Instituto de F\'isica Te\'orica - Universidade Estadual Paulista (UNESP)\\Rua Dr.~Bento Teobaldo Ferraz 271, 01140-070 S\~{a}o Paulo, SP Brazil}
\affiliation[b]{Centro de Investigaciones, Universidad Antonio Nari\~{n}o\\
Cra 3 Este \# 47A-15, Bogot\'{a}, Colombia}
\affiliation[c]{GRAPPA Institute, University of Amsterdam\\Science Park 904, 1090 GL Amsterdam, The Netherlands}
\affiliation[d]{LAPTh, CNRS, 9 Chemin de Bellevue, 74941 Annecy-le-Vieux, France}

\abstract{
Collider, space, and Earth based experiments are now able to probe several extensions of
the Standard Model of particle physics which provide viable dark matter candidates.
Direct and indirect dark matter searches 
rely on inputs of astrophysical nature, such as the
local dark matter density or the shape of the dark matter profile in the target in object.
The determination of these quantities is highly affected by astrophysical uncertainties.
The latter, especially those for our own Galaxy, are ill--known, and often not fully accounted for when analyzing the
phenomenology of particle physics models. 
In this paper we present a systematic, quantitative estimate of how
astrophysical uncertainties on Galactic quantities (such as the local
galactocentric distance, circular velocity, or the morphology of the stellar disk and bulge) 
propagate to the determination of the phenomenology
of particle physics models, 
thus eventually affecting the determination of new physics parameters.
We present results in the context of 
two specific extensions of the Standard Model
(the Singlet Scalar and the Inert Doublet)
that we adopt as case studies for their simplicity in illustrating the magnitude and impact of such uncertainties
on the parameter space of the particle physics model itself. 
Our findings point toward very relevant effects of current Galactic uncertainties
on the determination of particle physics parameters, 
and urge a systematic estimate of such uncertainties in more complex scenarios,
in order to achieve constraints on the determination of new physics that realistically 
include all known uncertainties.
}

\begin{document}
\begin{flushright}
PI/UAN-2016-598FT\\
LAPTH-071/16\\
\end{flushright}
\maketitle

\section{Introduction} 
\par Searches for the very nature of the elusive {\it dark} component of matter (DM) are experiencing a 
crucial moment in these very years: the enhanced sensitivity of direct and indirect searches, together
with the latest data coming from collider experiments, allows to constrain the parameter space of several
extensions of the Standard Model (SM) of particle physics, in some cases strongly challenging models which have been
very popular in the last years. The multichannel searches for DM are seeing the dawn of a real
precision era.
The grandeur of this endeavor carries the burden of precision, and it becomes timely and
mandatory to properly assess the entire budget of uncertainties that affect such an amazingly 
refined construction.
It is very well known, and we also recall in the following in more detail, that the particle 
interpretation of the data coming from direct and indirect searches depends on quantities of
astrophysical nature, such as the spatial distribution of  DM in the target for indirect searches, and its
phase space distribution in the solar neighborhood for direct ones.
Strenuous efforts are ongoing from the side of the astrophysical community to assess these quantities,
in a major endeavor involving astronomical observations of diverse nature. 
Yet, the determination of the quantities of interest are affected
by often sizable uncertainties.
This is also well known in the literature, where the entire extent of these uncertainties does not always
propagate its way in the determination of new physics.
\par In this work, we aim at presenting a case study by systematically analyzing how the
uncertainties on the DM structure in our Galaxy affect  the determination
of new physics. We will use two of the simplest possible extensions of the SM: the Singlet Scalar
(SSDM) 
and the Inert Doublet (IDM) 
DM models. Those models have been chosen as ideal testbeds given the relatively
simple dependence of their phenomenology on a limited set of parameters, which makes it easy to visualize
the effects of astrophysical (and in this case especially, Galactic) uncertainties in the parameter space
of the particle physics model itself.  
Our goal is to prompt the evidence for the relevance of the propagation
of known, but too often overlooked, unknowns of astrophysical nature directly into the determination of new physics.
In order to do so, we first present the most recent results on the determination of the DM
distribution in our Galaxy, and most relevantly its uncertainties. We then show the dependence of 
DM direct and indirect searches on the Galactic uncertainties, and specify how the constraints 
on the parameters of the IDM and SSDM models set by direct and indirect searches are affected by such uncertainties.

\par The paper has the following structure. In section \ref{sec:setups} we describe the setups of our analysis:
datasets and techniques adopted for the determination of the DM density structure of the Milky Way (MW); 
the adopted benchmark limits for DM direct detection;
and the probes chosen as indirect detection targets.
In section \ref{sec:parphysbench} we describe our benchmark
particle physics models: extensions of the SM which provide a well posed
DM candidate, and have recently been claimed to be strongly constrained by existing
data. 
In section \ref{sec:results} we present the finding of our analysis,
and the impact of uncertainties on Galactic morphologies and 
parameters on the determination of new physics. In our conclusions, we summarize our key results
and motivate how they prompt the extension of a similar complete analysis to
more complex extensions of the SM.

\vspace{0.5cm}

\section{Setups} 
\label{sec:setups}
\subsection{Galactic Dynamics}
\label{sec:astro}
\par In order to determine the DM density profile of our own Galaxy, we adopt a well known dynamical method: objects in circular orbits around the Galactic center (GC) are used as tracers of the total gravitational potential, and the rotation curve (RC) thus obtained (in the plane of the disk) is compared to the circular velocity expected to be caused by the visible component of the MW alone.
The mismatch between the two is accounted for by a non visible, {\it dark} component of matter, whose density distribution can be obtained by fitting an appropriately parametrized function to the total RC. This class of methods, often known as {\it global methods}, offers a series of advantages with respect to {\it local} ones, which permit to determine the DM distribution only in a small region around the location of the Sun, and a series of shortcomings. Both are carefully described in the recent review \cite{JustinRead2014}, and while addressing the reader to it and references therein for a complete overview, we highlight here the advantages and shortcomings of relevance to this specific analysis.
We follow the recent analysis in \cite{Pato:2015dua}, which respect to previous similar studies \cite{CatenaUllio2010,Iocco:2011jz,Nesti:2013uwa}, offers the remarkable improvement to adopt a vast range of data--driven morphologies for the three visible components of the MW (stellar bulge, stellar disk(s), gaseous disk). As shown therein, the choice of stellar bulge/disk affects the shape of the DM profile beyond the statistical uncertainties associated to each one of the visible components, leading to the conclusion that our ignorance on the morphology of the MW hinders our determination of the DM profile more than the uncertainties associated to their normalization. Also, a quantitative estimate of the effect of the currently quoted uncertainties on the Galactic parameters ($R_0$, $v_0$) is offered, showing effects comparable with that of the Galactic morphology.
This matter is certainly well known in principle, but again its actual magnitude is ill--known, and its effect on the determination of new physics is equally unaccounted for in most of the literature.
Unfortunately, none of these uncertainties are easily treatable in a statistical way, and one runs the risk to underestimate the effect of cross--correlations between datasets, or to be affected by hidden biases in the choice of the null hypothesis. 
For this reason, we aim here only to offer an estimate of the effects {\it in the parameter space of particle physics models}, by varying different sources of uncertainty one at a time.

\par By combining together one by one all possible combinations of bulge, disk, and gas, we obtain a set of unique ``{baryonic morphologies}'', i.e. a catalogue of
observationally--inferred morphologies, each one of them carrying a statistical uncertainty arising from the normalization of the density profile of each component, which is then propagated to the corresponding RC (generated by that specific configuration of visible matter) \cite{2015NatPh..11..245I, Pato:2015dua}. 
For each baryonic morphology and each set of Galactic parameters, separately, we add the RC due to DM to the one due to visible matter. The DM density profile is parameterised though a generalised NFW (gNFW) profile:
\begin{equation}
\rho_\text{DM} (R) = \rho_0 \left(\frac{R_0}{R}\right)^{\gamma}\left(\frac{R_s+R_0}{R_s+R}\right)^{3-\gamma} ,
\label{eq:gNFW}
\end{equation}
where $R_0$ is the distance of the Sun from the GC, $\rho_0$ the DM ``local'' density (i.e.~at the Solar position), 
$R_s$ the so-called ``scale radius'' of the DM density profile, and $\gamma$ the so-called ``profile index'' (the standard NFW profile having $\gamma$=1).
We compare the resulting  ``total'' (baryon + DM) RC to the latest compilation of observed RC data, presented in \cite{2015NatPh..11..245I}. 
Following closely the methodology of \cite{Pato:2015dua}, we scan the ($\rho_0$, $\gamma$) space, while keeping the scale radius $R_s$ constant.
We determine the goodness of fit of each point in the parameter space using the two--dimensional variable:

\begin{equation}
\chi^2 =\sum_{i=1}^N d_i^2\equiv \sum_{i=1}^N \left[\frac{(y_i-y_{t,i})^2}{\sigma_{y, i}^2+\sigma_{b,i}^2}+\frac{(x_i-x_{t,i})^2}{\sigma_{x,i}^2}\right] ,
\label{eq:chi2}
\end{equation}
where we have introduced the reduced variables \(x=R/R_0\) and \(y=w/w_0-1\); \((x_i\pm\sigma_{x,i},\,y_i\pm\sigma_{y,i})\) are the RC measurements, \(\sigma_{b, i}\) is the uncertainty of the individual baryonic model evaluated at \(x_i\) and \((x_{t, i},\,y_{t,i})\) are the points that minimize \(d_i\) along the curve \(y_t(x)=w_t(R=xR_0)/w_0-1\). The variable $w(R)=v_c(r)/R$ is the angular velocity at the galactocentric distance $R$ (with $w_0$=$w(R_0)$), and it is used as an independent variable as the uncertainties on $R$ and $w$ are uncorrelated (see \cite{2015NatPh..11..245I} and references therein).
The sum runs over all the $N$ objects in the compilation at \(R>R_{\rm cut}=2.5\) kpc in order to exclude the innermost regions of the Galaxy where axisymmetry breaks down and some tracers may present non-circular orbits.
The function in eq.~(\ref{eq:chi2}) has been shown to have a $\chi^2$ distribution for the case at hand in \cite{2015NatPh..11..245I}, and offers the advantage of an unbinned analysis which properly takes into account the statistical uncertainties of the observed RC dataset (in both dimensions), and that of the baryonic RC, propagated from the normalization of the stellar bulge and disk (respectively from microlensing optical depth in the direction of the bulge and local stellar surface density, see \cite{Iocco:2011jz, Pato:2015dua} for methodology, and references therein).
The ``best fit'' point is obtained by picking the point in ($\rho_0$, $\gamma$) space that minimizes the two dimensional $\chi^2$ described above, while we have kept the scale radius constant at the value $R_s$=20 kpc.
We note that also the variation of $R_s$ is expected to have some impact. Although 
we have tested that the choice of a fixed $R_s$ value does not affect significantly our conclusions,
a full analysis of the effect of the $R_s$ variation is beyond the scope of the present paper and we postpone it to a future work.~\footnote{We checked that
by varying its value by a factor 2 we observe a maximal variation in the local DM density of $\lesssim$5\%, and on the $\mathcal{J}$-factor (see section~\ref{sec:ID} for the definition of $\mathcal{J}$-factor) of $\lesssim$10\% for the region of interest of the GC GeV excess. As it will be seen, this is well below the effects of the variation of baryonic morphology or Galactic parameters.}

In order to probe the effect of different sources of ignorance, we test the following uncertainties one at a time:

\begin{itemize}
\item Statistical uncertainties;
\item Uncertainties on Galactic parameters, ($R_0$, $v_0$);
\item Uncertainties on the morphology of the visible, i.e.~baryonic, component of the Galaxy.
\end{itemize}

The numerical values adopted, the results obtained, as well as the reference for the morphologies that comply with the above conditions are presented schematically in table~\ref{tab:barbrack}, and we summarize here the criteria behind the choices adopted, according to the above rationale: 
\begin{itemize}
\item \par{\it Standard Galactic parameters.} The ``standard'' Galactic parameter values are ($R_0$, $v_0$)=(8 kpc, 230 km/s). When these values are adopted, the peak speed of the Maxwellian velocity distribution of DM particles is taken to be equal to the local circular speed, $v_{\rm peak}=v_0=230$ km$/$s.~\footnote{\label{ftn:2}This choice for the peak speed of the Maxwellian velocity distribution falls in the range of [223 -- 289] km$/$s suggested by high resolution hydrodynamic simulations~\cite{Bozorgnia:2016ogo} (see also section~\ref{sec:DD}). We have checked that varying $v_{\rm peak}$ does not make a visible difference in the direct detection limits in the parameter space of the SSDM and IDM models, and its effect is much smaller than the effect of variation of other Galactic parameters.}
\item \par{\it Reference morphology.} The ``representative'' baryonic morphology is \cite{Dwek1995, HanGould2003, Ferriere2007, Ferriere1998};
referred to as ``\refmorf{} '' in table~\ref{tab:barbrack}.
\item \par{\it Galactic parameters variation.} The extreme values for the Galactic parameters are chosen to vary between $R_0$=[7.5 -- 8.5] kpc and $v_0$=[180 -- 312] km/s, for our representative morphology \refmorf{}. The local circular speed can range from $(200 \pm  20)$~km/s to $(279 \pm 33)$ km/s~\cite{McMillan:2009yr}. Hence, we take $v_0=180$ km/s and 312 km/s as lower and higher estimates, respectively. When these values are adopted for $v_0$, we take $v_{\rm peak}=250$~km$/$s, regardless of $v_0$.$^\textrm{\ref{ftn:2}}$
\item \par{\it Morphology variation.} The extreme baryonic morphologies are chosen to be those that require the maximum/minimum values of $\gamma$, $\rho_0$, in order to visualize the maximum impact on both direct and indirect detection as we will see in the following sections.
\end{itemize}

With reference to the nomenclature in table~\ref{tab:barbrack}, we find that (for assigned, standard Galactic parameters) the baryonic morphologies that maximize/minimize
\begin{itemize}
\item the index $\gamma$ are respectively ``{\it FkX} '' and ``{\it DiX} '';
\item the local DM density $\rho_0$ are respectively ``{\it CjX} '' and ``{\it FiX} ''.
\end{itemize}

In figure~\ref{fig:barbrack} we display the rotation curves corresponding to the baryonic morphologies described above, assuming fixed Galactic parameters  ($R_0$, $v_0$)=(8 kpc, 230 km/s).
Statistical uncertainties associated to the displayed central values are not shown, but they are taken into account for the fitting procedure as described above.
We also display our compilation of data for the observed RC and their 1$\sigma$ uncertainties, as originally presented in \cite{2015NatPh..11..245I}.
In order to normalize the data to different values of the Galactic parameters, we have used the publicly available tool {\it galkin}, \cite{galkin}.
In table~\ref{tab:barbrack} we report the results of the fitting procedure described and the parameters of the selected 
morphologies.

When varying Galactic parameters, we obtain the total mass of the MW within 50 kpc to be in the range \(\mathrm{M(<50\;kpc)}=(1.2-22.9)\times 10^{11}\;\mathrm{M_{\odot}}\). 
The lower limit is in agreement with previous determinations~\cite{Wilkinson:1999hf, Sakamoto:2002zr, Smith:2006ym, Deason:2012wm, Gibbons:2014ewa}, 
while the larger MW masses we obtain can not be directly compared, as the the adopted Galactic Parameters are different than ours.
When varying baryonic morphology, the minimum/maximum values obtain are \(\mathrm{M(<50\,kpc)}=4.36^{+0.11}_{-0.10}\times 10^{11}\;\mathrm{M_{\odot}}\) and \(\mathrm{M(<50\,kpc)}=7.0\pm0.3\times 10^{11}\;\mathrm{M_{\odot}}\). The former value for the variation of morphology is in good agreement with mass estimate from kinematics of globular clusters, satellite galaxies and halo stars \cite{Wilkinson:1999hf, Sakamoto:2002zr,  Besla:2007kf, Deason:2012wm}. There is, however, a discrepancy at the \(1\sigma\) level with the independent determination in \cite{Gibbons:2014ewa}, that used the Sagittarius stream, and is slightly smaller
than the other cited determinations. 
All our results are in agreement at the \(1\sigma\) level with the recent estimate of the dynamical mass \cite{portail2015} within the region of the Galactic bulge, as 
in the analysis presented in \cite{Iocco:2016itg}.
 
In figure~\ref{fig:plotDens}, we show the DM density profiles corresponding to the selected morphologies 
in table~\ref{tab:barbrack}. When varying the morphology, almost all DM profiles are in agreement with recent 
findings for MW-like galaxies in hydrodynamical simulations~\cite{Calore:2015oya,Schaller:2015mua}.
The upper panel displays the DM density as a function of the distance from the 
GC, while in the lower panel the relative error with respect to the reference model is shown.
As a reference we also depict the traditionally adopted NFW profile, corresponding to a gNFW with parameters 
$\gamma = 1$, $\rho_0 = 0.4$ GeV/cm$^3$. 
\bgroup
\def\arraystretch{1.4}
\begin{table}
\resizebox{\textwidth}{!}{%
\begin{tabular}{| c || c | c | c | c | c | c | }
\hline
Morphology & \(R_0\) (kpc) & \(v_0\) (km/s) &  \(M_*\) (\(\times 10^{10}\,M_{\odot}\)) & \(\gamma\)  & \(\rho_0\) (GeV/\(\rm cm^3\)) & Reference \\
\hline \hline

BjX & 8 & 230 &   \(4.6^{+0.6}_{-0.5}\)  &  \(1.11^{+0.04}_{-0.03}\) & \(0.466 \pm 0.010\) & \cite{Dwek1995, HanGould2003, Ferriere2007, Ferriere1998} \\
\hline 
\hline
BjX & 7.5 & 312 & \(4.2^{+0.6}_{-0.5}\)  &  \(0.633^{+0.019}_{-0.020}\) & \(1.762\pm 0.017\) & \cite{Dwek1995, HanGould2003, Ferriere2007, Ferriere1998} \\
\hline
BjX  & 8.5 & 180 & \(5.1^{+0.7}_{-0.6}\)  &  \(2.02 \pm 0.07\) & \(0.055  \pm 0.004\) &  \cite{Dwek1995, HanGould2003, Ferriere2007, Ferriere1998} \\
\hline
\hline

FkX & 8 & 230 & \(4.3\pm0.5\) & \(1.38^{+0.03}_{-0.02}\) & \(0.427^{+0.007}_{-0.008}\)  & \cite{Robin2012, CalchiNovatiMancini2011, Ferriere2007, Ferriere1998} \\
\hline 
DiX  & 8 & 230 &  \(5.6\pm0.7\) & \(0.43 ^{+0.07}_{-0.06}\) & \(0.405 \pm 0.011\)  &  \cite{BissantzGerhard2002, Bovy:2013raa, Ferriere2007, Ferriere1998}\\
\hline
\hline 
CjX  & 8 & 230 & \(4.8^{+0.7}_{-0.6}\) &  \(1.03 \pm 0.04\) & \(0.471^{+0.010}_{-0.011}\)   &  \cite{Vanhollebeke2009, HanGould2003, Ferriere2007, Ferriere1998} \\
\hline
FiX  & 8 & 230 & \(5.2\pm0.6\)  & \(0.82\pm0.05\) & \(0.387\pm0.010\) & \cite{Robin2012,Bovy:2013raa,  Ferriere2007, Ferriere1998}  \\
\hline
\end{tabular}}
\caption{We adopt a gNFW density profile with \(R_s=20\) kpc. 
From left to right we report the nomenclature adopted for each morphology, the values of Galactic parameters ($R_0$, $v_0$), the
baryonic mass in the Galaxy for that specific baryonic morphology,  the best--fit values of index $\gamma$ and $\rho_0$ according to the procedure 
described in the text, and the references for the three-dimensional morphology shape. The criteria that led to the choice of these morphologies are explained in the text.}
\label{tab:barbrack}
\end{table}

\begin{figure}[t!]
	\centering
	\includegraphics[width=0.9\columnwidth]{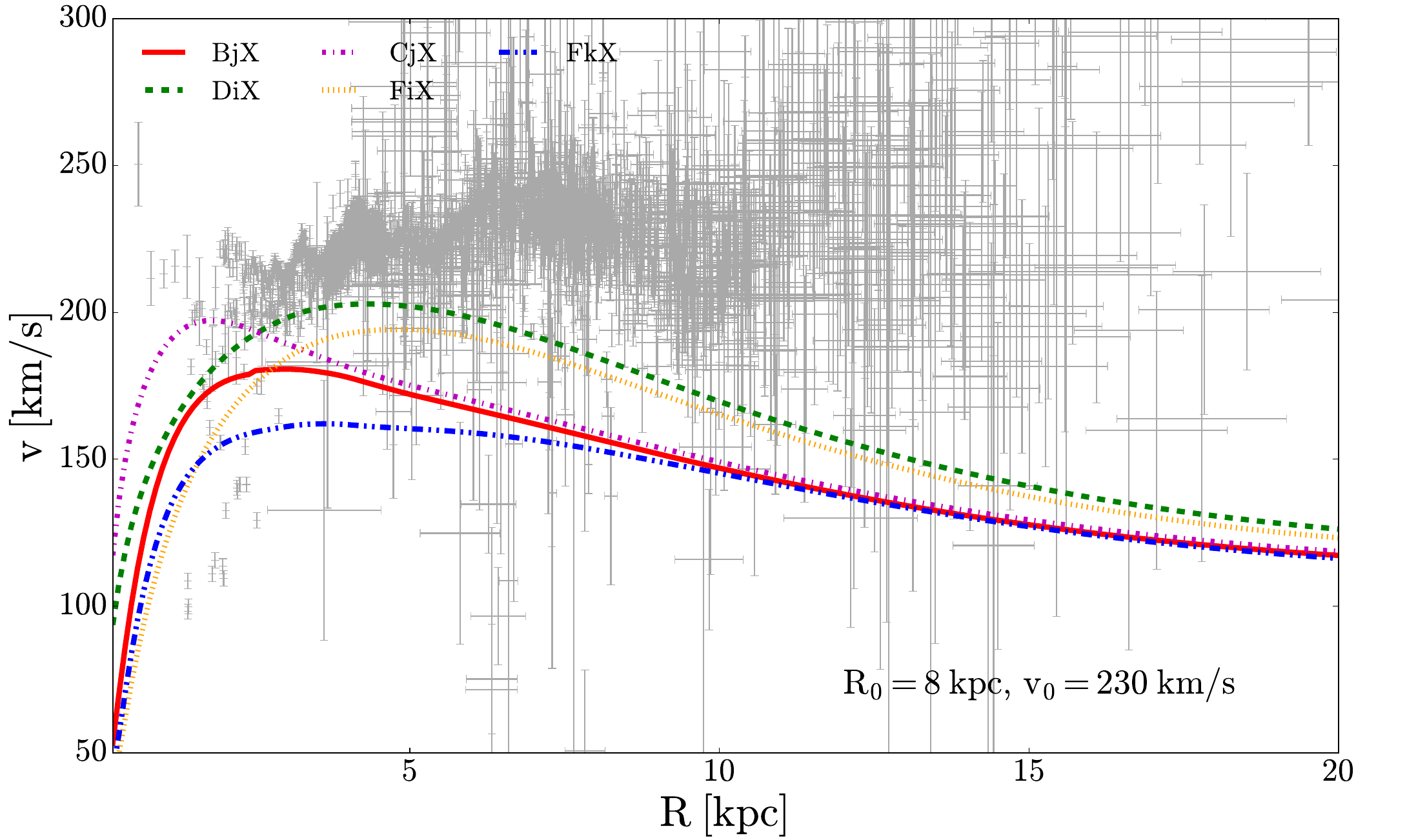}
\caption{RC produced by the benchmark baryonic morphologies (colored curves), reported in table~\ref{tab:barbrack} and described in the text. RC data points and their 1$\sigma$ errors shown in gray are from the compilation presented in \cite{2015NatPh..11..245I}.}
\label{fig:barbrack}
\end{figure}

\begin{figure}[t!]
	\centering
	\includegraphics[width=0.55\columnwidth]{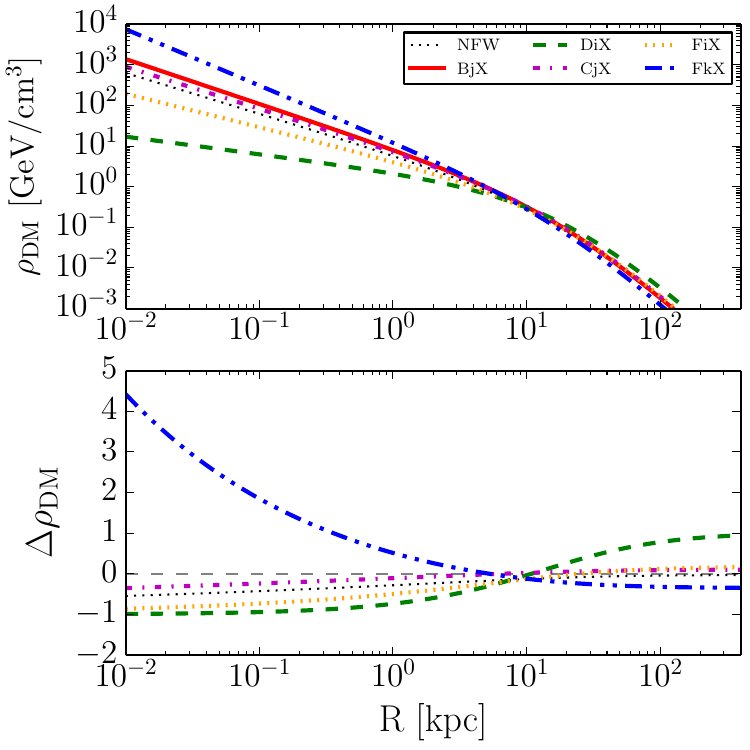}

\caption{
	The DM density, $\rho_{\rm DM}$, as a function of radial distance from the GC, $R$,
	for a standard NFW profile (black dotted line), and all baryonic morphologies in table~\ref{tab:barbrack}
	with standard Galactic parameters, as described in the text.
	The bottom panel refers to the relative error of $\rho_{\rm DM}$
	 with respect to the reference morphology  \refmorf{}. 
	 The dashed gray line in the bottom panel corresponds to a perfect match between 
	 the morphology considered and the  \refmorf{}. 
	 }
\label{fig:plotDens} 
\end{figure}


\subsection{Direct Detection}
\label{sec:DD}

The aim of direct detection experiments is to measure the small recoil energy of a nucleus in an underground detector after the collision with a WIMP arriving from the DM halo of the MW. The current status of direct detection searches is ambiguous with a few experiments reporting hints for a DM signal~\cite{Bernabei:2013xsa, Agnese:2013rvf, Aalseth:2010vx, Aalseth:2014eft}, while all other experiments report null-results. Currently the LUX (Large Underground Xenon) experiment~\cite{Akerib:2016vxi} places the strongest exclusion limit in the plane of spin-independent DM-nucleon cross section and WIMP mass for large DM masses, while the PandaX-II (Particle and Astrophysical Xenon Detector) experiment~\cite{Tan:2016zwf} has recently reported competitive null results. In this paper we focus on the impact of astrophysical uncertainties on the LUX exclusion limit in the parameter space of specific particle physics models. However, we note that the variation of the exclusion limits set by other direct detection experiments due to astrophysical uncertainties would be similar to those discussed for LUX.



\par For a DM particle scattering elastically off a nucleus with atomic mass number $A$, the differential event rate (per unit energy, per unit detector mass, per unit time) in direct detection experiments for the case of spin-independent scattering can be written as, 
\begin{equation}\label{eq:Reta}
\frac{d R}{d E_R} = \frac{\rho_0\, A^2\, \sigma_{\rm SI}}{2\, m_{\rm DM}\, \mu_{p}^2} \, F^2(E_R) \, \eta(v_{\rm min}, t),
\end{equation}
where $E_R$ is the nuclear recoil energy, $\rho_0$ is the local DM density, $m_{\rm DM}$ is the DM mass, $\mu_p$ is the reduced mass of the DM-nucleon system, $\sigma_{\rm SI}$ is the spin-independent DM-nucleon scattering cross section, and $F(E_R)$ is a form factor. $v_{\rm min}=\sqrt{m_A E_R/(2\mu_A^2)}$ is the minimum speed needed for the DM particle to deposit a recoil energy $E_R$ in the detector. Here $m_A$ is the mass of the nucleus, and $\mu_A$ is the DM-nucleus reduced mass. The halo integral, $\eta(v_{\rm min}, t)$, which together with the local DM density encompasses the astrophysics dependence of the recoil rate, is defined as,
\begin{equation}\label{eq:eta} 
\eta(v_{\rm min}, t) \equiv \int_{v > v_{\rm mim}} d^3 v \frac{f_{\rm det}({\bf v}, t)}{v} \,,
\end{equation}
where $f_{\rm det}({\bf v}, t)$ is the local DM velocity distribution in the detector rest frame. 

Eq.~\eqref{eq:Reta} can be written as,
\begin{equation}\label{eq:RetaFact}
\frac{d R}{d E_R} = C_{\rm PP} \, F^2(E_R) \, \rho_0 \, \eta(v_{\rm min}, t),
\end{equation}
where the coefficient $C_{\rm PP}=A^2\, \sigma_{\rm SI}/(2\, m_{\rm DM}\, \mu_{p}^2)$ contains the particle physics dependence of the event rate, while $\rho_0\, \eta (v_{\rm min}, t)$ contains the astrophysics dependence.

In the analysis of direct detection data, usually the Standard Halo Model (SHM) is adopted. In the SHM, the DM halo is spherical and isothermal, and the local DM velocity distribution is an isotropic Maxwell-Boltzmann distribution with a peak speed, $v_{\rm peak}$ equal to the local circular speed, $v_0$. 

The results of state-of-the-art high resolution cosmological simulations which include both DM and baryons indicate that a Maxwellian distribution with a best fit peak speed in the range of 223 -- 289 km$/$s fits well the local velocity distribution of simulated MW-like haloes~\cite{Bozorgnia:2016ogo}. Based on the results of ref.~\cite{Bozorgnia:2016ogo}, for the analysis of direct detection data in this work we adopt a Maxwellian velocity distribution truncated at the Galactic escape speed, and with a peak speed in the range of $[223 - 289]$~km$/$s, independent from the local circular speed. 
For the local circular speed, we adopt $v_0=180$~km$/$s and 312 km$/$s as high and low estimates. For the peculiar velocity of the Sun with respect to the Local Standard of Rest we assume $(11.10, 12.24, 7.25)$~km$/$s~\cite{Schoenrich:2009bx} in Galactic coordinates. We adopt the median value of the local Galactic escape speed reported by the RAVE survey, $v_{\rm esc}=533$~km$/$s~\cite{Piffl:2013mla}.

Recently, the LUX experiment has reported the results of 332 live days of data, with no evidence of a DM signal~\cite{Akerib:2016vxi}. Since the exposure and detector response information is not publicly available for each event in the recent LUX data, we perform an analysis of the 2015 LUX results~\cite{Akerib:2015rjg} instead. In ref.~\cite{Akerib:2015rjg}, the LUX collaboration presented an improved analysis of their 2013 data for an exposure of $1.4 \times 10^4$ kg days. To set an exclusion limit using the LUX data, we employ the maximum gap method~\cite{Yellin:2002xd}, since we cannot reproduce the likelihood analysis performed by the LUX collaboration with the available information. We consider the events which fall below the mean of the nuclear recoil band (red solid curve in figure~2 of \cite{Akerib:2015rjg}) as signal events, and assume an additional acceptance of 0.5. As seen from figure~2 of \cite{Akerib:2015rjg}, one event makes the cut. We take the detection efficiency from figure~1 of \cite{Akerib:2015rjg}, and set it equal to zero below the recoil energy of $E_R=1.1$ keV, following the collaboration. Since we are only interested in events at $<18$ cm radius, we multiply the efficiency by $(18/20)^2$. To find the relation between $E_R$ and the primary scintillation signal S1, we find the value of S1 at the intersection of each recoil energy contour and the mean nuclear recoil curve from figure~2 of \cite{Akerib:2015rjg}. Assuming a Maxwellian velocity distribution with the same parameters as in \cite{Akerib:2015rjg}, we can find an exclusion limit at 90\% CL which agrees well with the limit set by the LUX collaboration.

In figure~\ref{fig:LUX-msigma}, we show the LUX exclusion limit in the $m_{\rm DM}-\sigma_{\rm SI}$ plane for the standard choice of Galactic parameters $(R_0, v_0)$ and the local DM density $\rho_0$ for our reference morphology given in the first row of table~\ref{tab:barbrack} (``reference model'' in figure legend), as well as for two representative variations of Galactic parameters and $\rho_0$ given in the second and third rows of table~\ref{tab:barbrack}. The largest variation of the exclusion limit with respect to the reference limit is due to the variation in the local DM density. Notice that in the exclusion limits shown in figure~\ref{fig:LUX-msigma}, we take $v_{\rm peak}=v_0$, while in the figures of section~\ref{sec:results} where we vary the Galactic parameters, we adopt a peak speed value independent of the local circular speed.
\begin{figure}[t!]
\centering
\includegraphics[width=0.49\columnwidth]{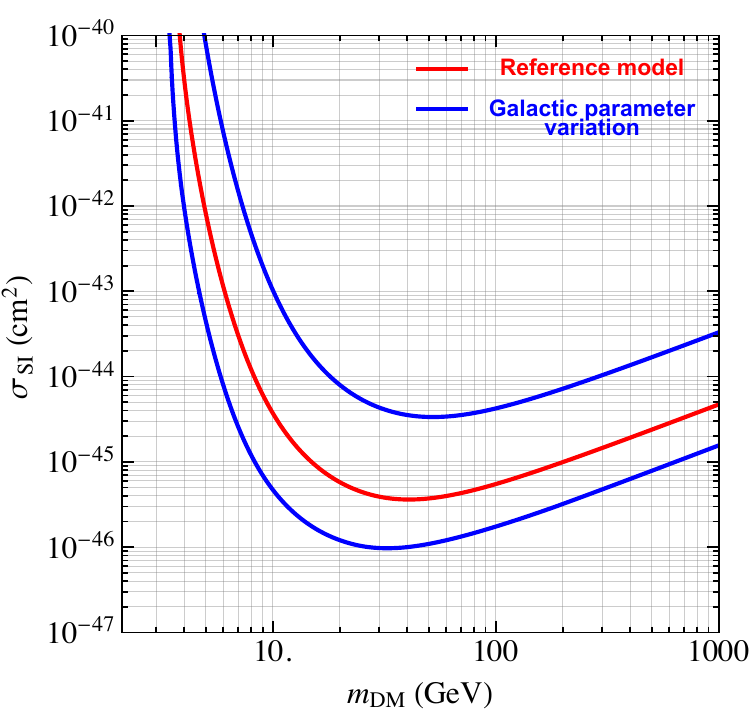} 
\caption{LUX exclusion limit at 90\% CL in the spin-independent DM-nucleon cross section and DM mass plane for the reference choice of Galactic parameters and local DM density (red curve) given in the first row of table~\ref{tab:barbrack}, and two variations of Galactic parameters and local DM density (blue curves) given in the second and third rows of table~\ref{tab:barbrack}, see text in section~\ref{sec:astro} for a complete description.}
\label{fig:LUX-msigma} 
\end{figure}

\subsection{Indirect Detection}
\label{sec:ID}
\par Indirect detection aims at detecting the flux of final stable particles
produced by DM annihilation or decay. Among those, gamma rays are considered the golden channel
for the identification of a possible DM signal since they preserve the spectral and spatial information of the 
signal itself. 
In the present work, we focus on the gamma-ray flux from DM annihilation for a twofold reason:
First, we work with gamma rays since the expected flux is directly expressed in terms of the DM density distribution. For charged cosmic rays, instead, the expected flux at Earth is obtained after propagating 
the produced particles in the interstellar medium and thus the dependence of the propagated flux on the 
DM density is less trivial.\footnote{We note however that the  ``source term'' of 
 charged particles produced by DM depends directly on the DM density and
thus is affected by the same uncertainties that we discuss explicitly for gamma rays.}
Secondly, the choice of the annihilation process -- instead of the decay one -- is motivated by the fact that 
the predicted flux is more affected by the astrophysical uncertainties on the 
DM density, since it depends on the DM density squared. 
We stress that our aim is to give concrete -- and intuitive --  examples of 
the effect of the astrophysical uncertainties for DM phenomenology.  

Typically, the gamma-ray flux from DM particles annihilation or decay can be factorized in terms of a 
particle physics term, $\Phi_{\rm PP}$, which contains the information of the underlying particle
physics theory of DM, and an astrophysical term, $\mathcal{J}$, which instead encodes 
the information about the geometrical distribution of DM in space. The DM expected flux writes as:
\begin{equation}
\Phi_{\rm DM}(E) =  \Phi_{\rm PP}(E) \; \mathcal{J} \,.
\label{eq:phflux}
\end{equation}
In this section, we remain agnostic about the particle physics sector, while we are interested
in quantifying the uncertainty affecting the \Jf~in light of the analysis performed in section~\ref{sec:astro}.
The \Jf~is defined as the integral along the line-of-sight of the DM density, $\rho_{\rm DM}$, in case of 
DM decay, or of the DM density squared,  $\rho_{\rm DM}^2$, in case of DM annihilation.
While the uncertainty of $\rho_{\rm DM}$ translates linearly in the uncertainty on $\mathcal{J}_{\rm decay}$, 
the $\mathcal{J}_{\rm annih}$ is more sensitive to the uncertainty on $\rho_{\rm DM}$ given the squared dependence.
In the case of DM annihilation: 
\begin{equation}
\label{eq:geofactor}
\mathcal{J} _{\rm annih} = \int_{\rm{l.o.s}} \rho_{\rm DM}^2(R(l,\psi)) dl ,
\end{equation}
where $\psi$ is the opening angle between the line of sight $l$ and the direction towards the GC.
The radial distance from the GC is $R^2 = R_0^2 + l^2 -2\,l\,R_0\,\cos{(\psi)}$.

\smallskip

In figure~\ref{fig:plotID} we show the uncertainty on the 
DM annihilation \Jf~(bottom panel) for different morphologies as in figure~\ref{fig:plotDens} (cf.~table~\ref{tab:barbrack}).
In the upper panel, we show the \Jf~as a function of the angle $\psi$, comparing a standard NFW profile
with our reference baryonic morphology and other morphology configurations, cf.~table~\ref{tab:barbrack}.
In the inner region, i.e. within few degrees, the astrophysical uncertainty on the \Jf~is $\gtrsim \mathcal{O}$(10).

\smallskip

Accounting for the astrophysical uncertainty on the predicted DM flux is crucial
when comparing results from different targets.
For example, a positive signal might be seen in a gamma-ray target and interpreted in terms
of DM annihilation. The preferred particle physics parameter space, typically the average velocity annihilation cross section
$\sv$ vs DM mass $m_{\rm DM}$, depends on the \Jf~assumed for the target considered. 
On the other hand, null results from other targets impose upper limits on the allowed
($\sv$, $m_{\rm DM}$) parameter space. It might occur that the constraints are in tension with the signal.
However, such a tension relies on the assumed \Jf~and, thus, the uncertainty on the \Jf~must
 be fully accounted for before claiming a strong tension.
This is what happened for example in the case of the GC GeV excess (see e.g.~\cite{Calore:2014xka,TheFermi-LAT:2015kwa}): the DM
interpretation of such an excess started to be challenged by 
the latest constraints from dwarf spheroidal galaxies (dSphs)~\cite{Fermi-LAT:2016uux}.

Here we demonstrate that such a tension can be alleviated (or worsened) by the 
unavoidable uncertainty on the
\Jf~of the MW -- we do not discuss here the uncertainties on the dSphs \Jf~related to the choice of the dSphs DM profile (see e.g.~\cite{Strigari:2013iaa} for a discussion), 
nor the possible effect of varying the Sun position, $R_0$, on the stacked dSphs limit.\footnote{For a single dSph, under the assumption of point-like source emission, 
$\mathcal{J}\propto 1/d^2$, being $d$ the distance from the observer and, thus, depending on the Sun position, $R_0$, in the Galactic reference frame. 
In the simplest case of a dSph located at the GC: $\mathcal{J}(R_0^a)/\mathcal{J}(R_0^b) \propto (R_0^b/R_0^a)^2$. However, 
when considering more realistic geometries and the stacking of the dSphs, as done to derive the limits in ref.~\cite{Fermi-LAT:2016uux}, the dependence from $R_0$ is not anymore trivial and 
assessing its effect would require a re-analysis of the dSphs data which is beyond the scope of the present paper.}
In figure~\ref{fig:GCvsdwarfs}, we show the latest dSphs limits~\cite{Fermi-LAT:2016uux} and the region preferred by the
GC GeV excess for DM annihilating into a pair of b-quarks~\cite{Calore:2014nla}, 
for our reference morphology \refmorf{}
with standard Galactic parameters (``reference model'' in figure legend), and for the variation of Galactic parameters, as given in the second and third rows of table~\ref{tab:barbrack}.
For the sake of clarity we show only the 1 and 2$\sigma$ contours rescaled to the morphology of interest from the 
contours in ref.~\cite{Calore:2014nla}. The rescaling of the original contours (in figure~\ref{fig:GCvsdwarfs} as well as in all other figures showing the GeV excess rescaled contours) 
 is made by imposing that the GeV excess flux 
measured at 2 GeV and 5$^\circ$ away from the GC is conserved, in analogy to what done in ref.~\cite{Calore:2014nla}. Indeed, 
at 5$^\circ$ away from the GC the GeV excess intensity has been shown to be almost independent on the assumption on the gNFW slope used in the data analysis.
Therefore, the rescaling of the contours only involves $\mathcal{J}$($\psi = 5^\circ$), and not the integral over the full region-of-interest.
The variation of the contours in $\sv$ due to the uncertainty on Galactic parameters ($R_0$, $v_0$) can be larger than a factor of two in both directions.
We emphasise that also constraints on DM from the annihilation in the MW halo or 
the GC region would be affected by an analogous uncertainty.

\begin{figure}[t!]
	\centering
	\includegraphics[width=0.55\columnwidth]{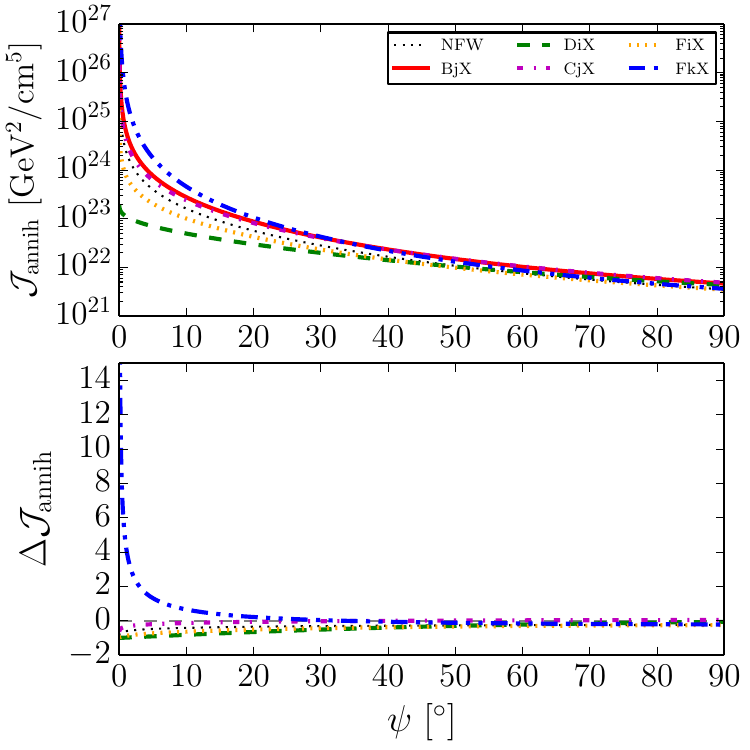} 
\caption{
	The DM annihilation \Jf~as a function of the angle $\psi$ between the line-of-sight and the GC
	for the same DM profiles in table~\ref{tab:barbrack} and figure~\ref{fig:plotDens} (same colors).
	The bottom panel refers to the relative error of the \Jf~with respect to the reference morphology \refmorf{}.
	The dotted line corresponds to the traditional NFW profile.
	The dashed gray line in the bottom panels corresponds to a perfect match between 
	 the morphology considered and  \refmorf{}. }
\label{fig:plotID} 
\end{figure}

\begin{figure}[t!]
	\centering
	\includegraphics[width=0.49\columnwidth]{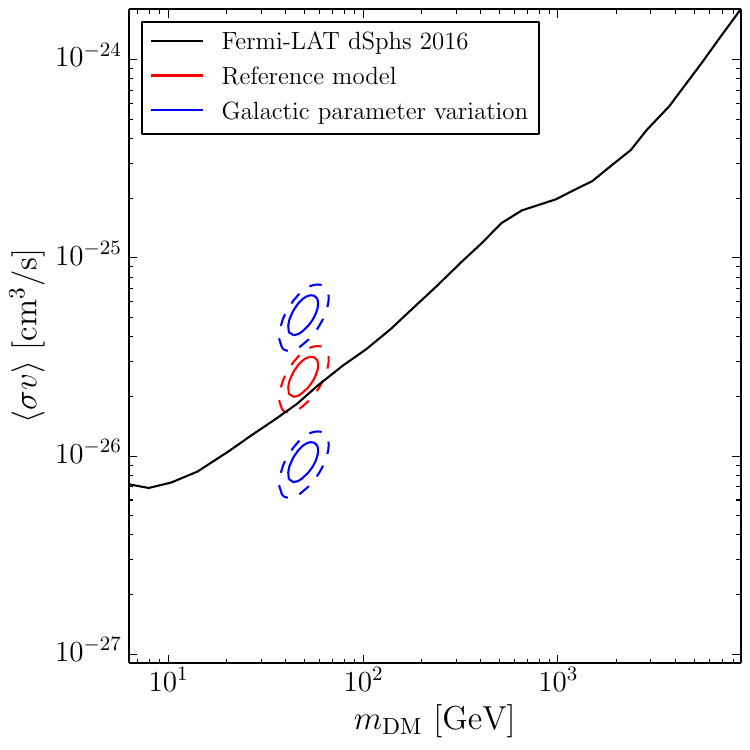}
\caption{Upper limit on DM annihilation cross section (into $\bar{b}b$), $\sv$, vs DM mass, 
	$m_{\rm DM}$, from the analysis of gamma rays from dwarf spheroidal galaxies~\cite{Fermi-LAT:2016uux} (black line).
	Best fit contours at 2$\sigma$ of the GeV excess as due to DM annihilation in b-quark pairs 
	for the gNFW parameters preferred by our reference morphology \refmorf{} (red contours)
	with standard Galactic parameters, and the same morphology by varying Galactic parameters 
	(blue countours), as given in the second and third rows of table~\ref{tab:barbrack}, 
	see text in section~\ref{sec:astro} for a complete description. This setup is the same as 
	used for the variations in figure~\ref{fig:LUX-msigma}.}
\label{fig:GCvsdwarfs} 
\end{figure}

\section{Particle Physics Benchmarks}
\label{sec:parphysbench}
\par We now apply the setup shown until now to some benchmark DM particle models,
in order to show how the astrophysical uncertainties affect the determination of
the physical parameters of the specific models at hand, in the context of DM direct and indirect detection.
We concentrate on two minimal extensions of the SM, 
the SSDM and the IDM, as they are arguably among the 
most minimal models for which it is easy to quantify
and visualize in a clear way the effects described above.
We do expect that our results will prompt a generalization to more complex scenarios
in which the effects are not so trivially discernible from effects due to the interplay
of numerous model parameters. In this section we present the general framework of these two simple
models and depict the state-of-the-art constraints on the model parameter space. The discussion
of the impact of astrophysical uncertainties will be the focus of section~\ref{sec:results}.

\subsection{Singlet Scalar Model}\label{sec:singlet}
The SSDM~\cite{McDonald:1993ex,Burgess:2000yq} is one of the minimal extensions of the SM that can provide a viable DM particle candidate. In addition to the SM, this model contains a real scalar $S$, singlet under the SM gauge group, but odd under a $\mathbb{Z}_2$ symmetry in order to guarantee its stability. 
The most general renormalizable scalar potential is given by
\begin{equation}
V=\mu_H^2\,|H|^2+\lambda_H\,|H|^4+\mu_S^2\,S^2+\lambda_S\,S^4+\lambda_{HS}\,|H|^2\,S^2\,,
\end{equation}
where $H$ is the SM Higgs doublet.
It is required that the Higgs gets a non-vanishing vacuum expectation value, $v_H = 246$~GeV, while the singlet does not, $\langle S\rangle = 0$.
At tree level, the singlet mass is $m_S^2=2\,\mu_S^2+\lambda_{HS}\,v_H^2$.
The phenomenology of this model is completely determined by three parameters: the DM mass $m_S$, the Higgs portal coupling $\lambda_{HS}$ and the quartic coupling $\lambda_S$.
Note however that $\lambda_S$ plays a minor role in vanilla WIMP DM phenomenology,~\footnote{Whereas $\lambda_S$ has a crucial role in DM phenomenology, in scenarios where DM is a SIMP, with sizable self-interactions~\cite{Bernal:2015xba}.} and thus hereafter we will focus only on the parameters $m_S$ and $\lambda_{HS}$.
These two parameters are the ones that determine the strength of both the direct and the indirect detection signals.

There has been a large amount of research on the SSDM, most of it focused on the WIMP scenario, where the singlet $S$ mixes relatively strongly with the Higgs and undergoes a thermal freeze-out. This scenario has been highly constrained by collider searches~\cite{Barger:2007im,Djouadi:2011aa,Djouadi:2012zc,Damgaard:2013kva,No:2013wsa,Robens:2015gla}, DM direct detection~\cite{He:2009yd,Baek:2014jga, Feng:2014vea, Han:2015hda} and indirect detection~\cite{Yaguna:2008hd, Goudelis:2009zz, Profumo:2010kp,Cline:2013gha,Urbano:2014hda, Duerr:2015mva,Duerr:2015aka}. 

\par We show the current constraints in figure~\ref{fig:SSDM}.
In both panels, the black thick line corresponds to the points that generate a DM relic abundance in agreement with the measurements by Planck~\cite{Ade:2015xua}, and the gray region below the line is excluded because it produces a too large DM abundance, thus overclosing the Universe.
The hatched light blue region in the upper left corner is forbidden by current constraints on the strength of the Higgs portal.
Indeed, for $m_S<m_{h^0}/2\sim62$~GeV, the Higgs can decay into a pair of DM particles, thus the current limits on the invisible Higgs branching ratio (BR$_\text{inv}\lesssim 20\%$~\cite{Bechtle:2014ewa}) and the Higgs total decay width ($\Gamma_{h^0}^\text{tot}\lesssim 22$~MeV~\cite{Khachatryan:2014iha}) constrain the Higgs coupling with the dark sector, $\lambda_{HS}$.

\par In the left panel of figure~\ref{fig:SSDM}, we display the exclusion limit on the spin-independent elastic WIMP-nucleon cross section at 90\% CL from the 2015 LUX results~\cite{Akerib:2015rjg}, which is translated into the dark red region in the top part of the figure. We plot here the limit derived from the red curve in figure~\ref{fig:LUX-msigma} described in section \ref{sec:DD}. We recall that this limit has been derived assuming the parameters in the first row of table~\ref{tab:barbrack}, i.e. a Maxwellian velocity distribution with $v_{\rm peak} = v_0 = 230$~km/s, $v_\text{esc} = 533$~ km/s, and $\rho_0 = 0.466$~ GeV/cm$^3$. In the right panel, we show instead the current limits from the analysis of dwarf spheroidal galaxies (dSphs) with
{\it Fermi}-LAT~\cite{Fermi-LAT:2016uux}. The region in blue represents the parameter space favoured by the interpretation of the GC excess (at 2$\sigma$), and corresponds to the red contour in figure~\ref{fig:GCvsdwarfs}, as described in section \ref{sec:ID}.

\begin{figure}[t!]
\centering
\includegraphics[width=0.45\columnwidth]{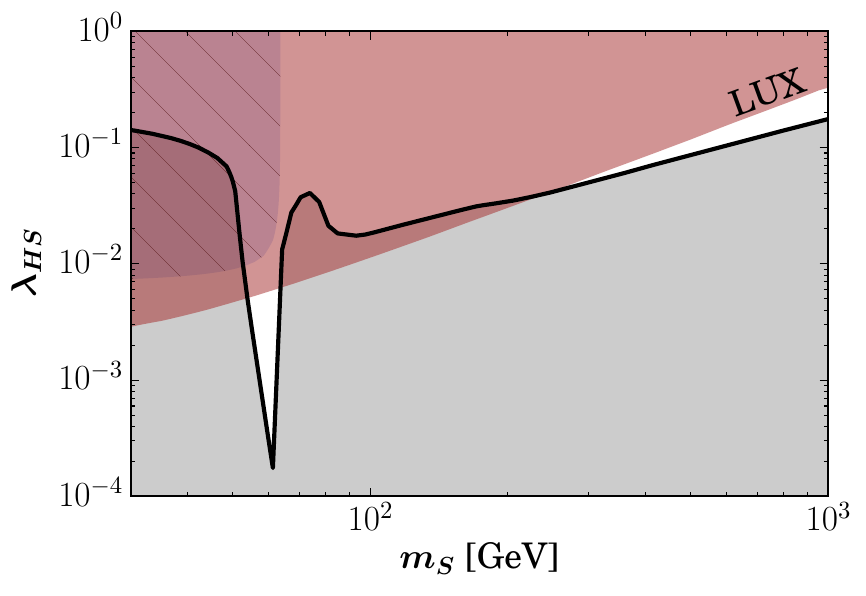}
\includegraphics[width=0.45\columnwidth]{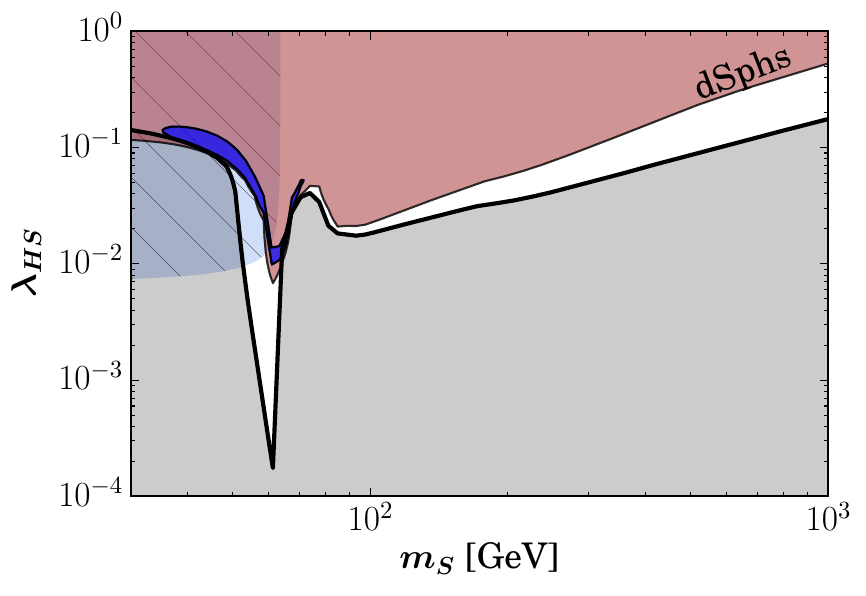}
\caption{\emph{Singlet Scalar Model}. In both panels, the black line corresponds to the points that 
generate a DM relic abundance in accordance to the measurements by Planck~\cite{Ade:2015xua}; 
the lower gray region over-closes the Universe. The upper left region (hatched light blue) 
is ruled out by the invisible decay of the Higgs~\cite{Bechtle:2014ewa, Khachatryan:2014iha}.
The upper dark red region in the left panel corresponds to the LUX exclusion limit on the spin-independent elastic WIMP-nucleon cross section at 90\% CL for the choice of parameters in the first row of table~\ref{tab:barbrack}, while the one in the right panel is derived from the limits on the averaged velocity annihilation cross section from the combined analysis of 
dwarf spheroidal galaxies in the MW~\cite{Fermi-LAT:2016uux}.
The parameter space favoured by the GeV excess data~\cite{Calore:2014nla} (at 2$\sigma$)
is depicted by the blue region in the right panel.
}
\label{fig:SSDM} 
\end{figure}

\par From figure~\ref{fig:LUX-msigma}, we see that the LUX limit strongly depends on the astrophysical uncertainties on the Galactic parameters, and especially on the uncertainty in the local DM density.
Therefore, the available parameter space of the SSDM will depend on the actual 
configuration of Galactic parameters. 
On the other hand, in the case of indirect searches, we do not explore uncertainties on the limits imposed by dSphs, 
but we investigate how the region favoured by the DM interpretation of the GeV excess will move because of Galactic uncertainties, as already shown in figure~\ref{fig:GCvsdwarfs}.
We will show the response of the constraints to astrophysical uncertainties, and its implications in section \ref{sec:results}.

\subsection{Inert Doublet Model}\label{sec:doublet}
The IDM~\cite{Barbieri:2006dq} is another minimal extension of the SM that contains a second complex scalar doublet. The model contains an exact $\mathbb{Z}_2$ symmetry under which all SM particles --including one of the scalar doublets-- are even, and the second scalar doublet is odd. Since this discrete symmetry prevents mixing between the scalars, one of the doublets ($H$) is identified with the SM Higgs doublet. The second doublet ($\Phi$), odd under the $\mathbb{Z}_2$ parity, is inert in the sense that it does not couple to the SM particles.
The most general renormalizable scalar potential of the IDM is given by
\begin{equation}
V = \mu_1^2 |H|^2  + \mu_2^2|\Phi|^2 + \lambda_1 |H|^4+ \lambda_2 |\Phi|^4 + \lambda_3 |H|^2| \Phi|^2
                + \lambda_4 |H^\dagger\Phi|^2 + \frac{\lambda_5}{2} \Bigl[ (H^\dagger\Phi)^2 + \mathrm{h.c.} \Bigr]\,.
\label{Eq:TreePotential}
\end{equation}
In the general case, the $\lambda_i$ are complex parameters. Although considering this possibility can have interesting consequences for CP-violation and electroweak baryogenesis \cite{Chowdhury:2011ga, Borah:2012pu, Cline:2013bln}, in this work we limit ourselves to the case of real values. Upon electroweak symmetry breaking, the two doublets can be expanded in components as
\begin{equation}
        H ~=~ \left( \begin{array}{c} 0 \\ \frac{1}{\sqrt{2}}\left(v_H+h^0\right) \end{array} \right),
        \qquad
        \Phi ~=~ \left( \begin{array}{c} H^+\\ \frac{1}{\sqrt{2}}\left(H^0+\mathrm{i}A^0\right) \end{array} \right)\,.
\end{equation}
The $h^0$ state corresponds to the physical SM-like Higgs-boson.
The inert sector consists of a neutral CP-even scalar $H^0$, a pseudo-scalar $A^0$, and a pair of charged scalars $H^\pm$.
The $\mathbb{Z}_2$ symmetry guarantees the stability of the lightest state of the dark sector.
If it is neutral (either $H^0$ or $A^0$), this state can play the role of the DM.

At the tree level, the scalar masses are 
\begin{align}
m_{h^0}^2 &= \mu_1^2 + 3 \lambda_1\,v_H^2\,, \\
m_{H^0}^2 &= \mu_2^2 + \lambda_L\, v_H^2\,, \label{Eq:mH0tree} \\
m_{A^0}^2 &= \mu_2^2 + \lambda_S\, v_H^2\,, \\
m_{H^\pm}^2 &= \mu_2^2 + \frac{1}{2} \lambda_3\, v_H^2\,,
\end{align}
where $\lambda_L \equiv \frac{1}{2} \left( \lambda_3 + \lambda_4 + \lambda_5 \right)$ and $\lambda_S\equiv \frac{1}{2} \left( \lambda_3 + \lambda_4 - \lambda_5 \right)$.
The IDM scalar sector can be fully specified by a total of five parameters: three masses ($m_{H^0}$, $m_{A^0}$ and $m_{H^\pm}$) and two couplings ($\lambda_L$ and $\lambda_2$).
However, in this analysis the role of $\lambda_2$ will be disregarded, as it appears only in quartic self couplings among dark particles and does therefore not enter in any physically observable process at the tree level.~\footnote{Loop corrections to the WIMP DM phenomenology in the IDM have been studied in ref.~\cite{Goudelis:2013uca}.}

The phenomenology of the IDM has been largely studied since the model allows to generate a population of WIMP DM particles in the early Universe via a thermal freeze-out and it induces potentially observable signals in direct and indirect DM searches~\cite{Barbieri:2006dq,Majumdar:2006nt,LopezHonorez:2006gr,Gustafsson:2007pc,Agrawal:2008xz,Andreas:2009hj,Nezri:2009jd,Arina:2009um,Honorez:2010re,LopezHonorez:2010tb,Gustafsson:2010zz,Majumdar:2006nt,Klasen:2013btp,Garcia-Cely:2013zga,Garcia-Cely:2015khw}, collider searches~\cite{Barbieri:2006dq,Cao:2007rm,Lundstrom:2008ai,Dolle:2009ft,Miao:2010rg,Arhrib:2012ia,Swiezewska:2012eh,Krawczyk:2013jta, Goudelis:2013uca} and electroweak precision tests~\cite{Barbieri:2006dq,Grimus:2008nb}.

For this analysis, we perform a numerical analysis scanning randomly over $10$~GeV$<m_{H^0}<1$~TeV, $10^{-5}<|\lambda_L|<10^0$ and over $m_{A^0}$ and $m_{H^\pm}$ in the range [$m_{H^0},\,10$~TeV].
A number of theoretical and experimental constraints, largely discussed in the literature, can be imposed on the parameters of the model like perturbativity, vacuum stability~\cite{Gunion:2002zf,Gustafsson:2010zz,Khan:2015ipa}, unitarity of the S-matrix~\cite{Ginzburg:2004vp,Branco:2011iw}, electroweak precision tests~\cite{Arhrib:2012ia} and collider searches from LEP~\cite{Pierce:2007ut,Lundstrom:2008ai} and LHC~\cite{Arhrib:2012ia,Swiezewska:2012eh,Krawczyk:2013jta,Goudelis:2013uca,Pierce:2007ut}.
We restrict ourselves to the points fulfilling the observed DM relic abundance and the previously mentioned 
direct and indirect detection constraints.

\par Figure~\ref{fig:IDM} shows the current constraints on the IDM plane $(|\lambda_L|,\,m_{H^0})$.
In both panels, the visible dots correspond to those points of the parameter space that generates a DM relic abundance in agreement with Planck~\cite{Ade:2015xua}.
The light gray points in the upper left corner, however, are ruled out because they give rise to a too large invisible Higgs decay, in tension with LHC measurements~\cite{Bechtle:2014ewa,Khachatryan:2014iha}.
The left panel also displays the exclusion limit on the spin-independent elastic WIMP-nucleon cross section at 90\% CL from the 2015 LUX results~\cite{Akerib:2015rjg} for the choice of parameters in the first row of table~\ref{tab:barbrack} -- as described in section~\ref{sec:DD}, and shown as the red curve in figure~\ref{fig:LUX-msigma} -- which is translated into the dark red region in the top part of the figure.
In the right panel, we show instead the points (light brown) in tension with the analysis of dSphs with {\it Fermi}-LAT~\cite{Fermi-LAT:2016uux}:
the region in blue represents the parameter space favoured by the interpretation of the GC excess (at 2$\sigma$), and corresponds to the red contour in figure~\ref{fig:GCvsdwarfs}, as described in section~\ref{sec:ID}.
Let us emphasize that, as seen in figure~\ref{fig:GCvsdwarfs}, in this reference model the GC excess is in tension with the measurements of dSphs.

\begin{figure}[t!]
\centering
\includegraphics[width=0.45\columnwidth]{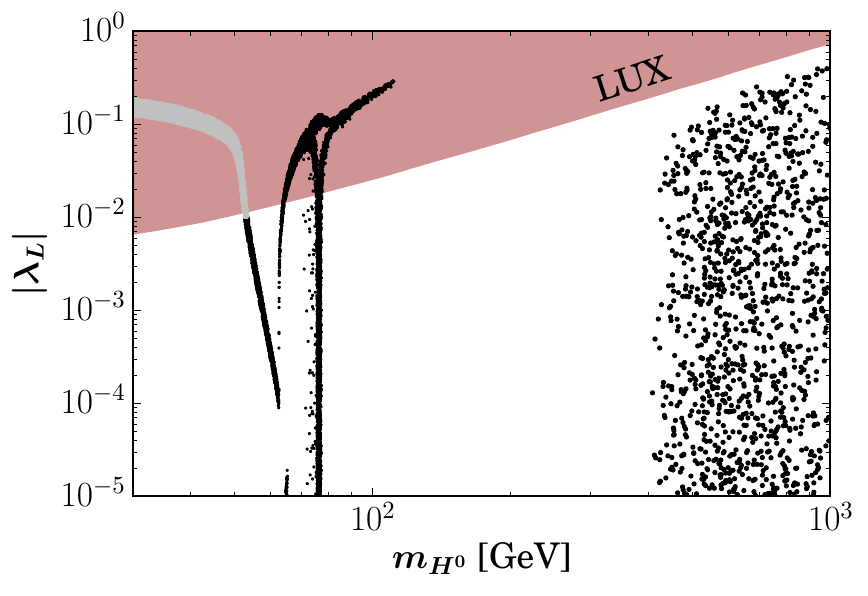}
\includegraphics[width=0.45\columnwidth]{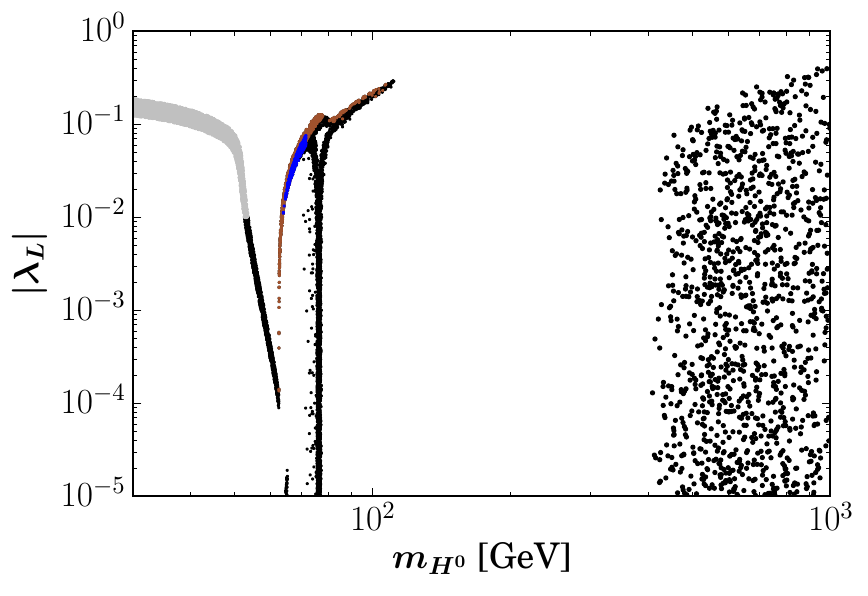}
\caption{\emph{Inert Doublet Model}.
The dots correspond to the parameter space that generates the correct DM relic abundance.
The light gray points give rise to a too large invisible Higgs decay, in tension with LHC measurements.
The dark red region in the left panel displays the LUX exclusion limit on the spin-independent elastic WIMP-nucleon cross section at 90\% CL for the choice of parameters in the first row of table~\ref{tab:barbrack}. In the right panel, we show, in blue, the points of the parameter space which can successfully explain the GeV excess (at 2$\sigma$), 
corresponding to the red contour in figure~\ref{fig:GCvsdwarfs}.
The points in light brown are those in tension with the analysis of dSphs with {\it Fermi}-LAT~\cite{Fermi-LAT:2016uux}.
}
\label{fig:IDM} 
\end{figure}

\par In figure~\ref{fig:IDM} we study the constraints on the IDM parameter space coming from a specific Galactic model (i.e. the reference morphology model  \refmorf{}).
In section~\ref{sec:results} we show the response of the constraints to astrophysical uncertainties, and its implications for both direct and indirect DM searches.

\section{Results}
\label{sec:results}
\par With all elements at hand, we now turn to show the effect of astrophysical uncertainties 
directly onto the parameter space of particle physics models discussed in the previous section.

\subsection{Impact on SSDM Parameter Determination}

\par We start by comparing the limit imposed by direct detection on the SSDM for different cases of variation of astrophysical uncertainties: In figure~\ref{fig:SSDM-1} we show how the LUX exclusion limit shown in the left panel of figure \ref{fig:SSDM} varies by including uncertainties arising from
{\it a)} the statistical uncertainty on our reference morphology, {\it b)} the variation of the Galactic parameters for the reference morphology, and {\it c)} the baryonic morphologies that maximize/minimize the local DM density $\rho_0$, as discussed in section~\ref{sec:astro}.

\begin{figure}[t!]
\centering
\includegraphics[width=0.32\columnwidth]{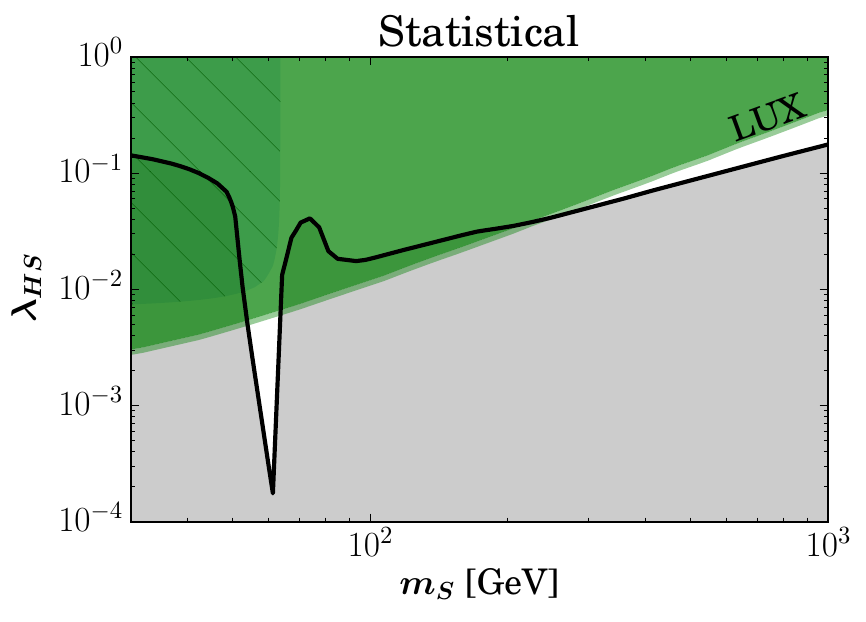}
\includegraphics[width=0.32\columnwidth]{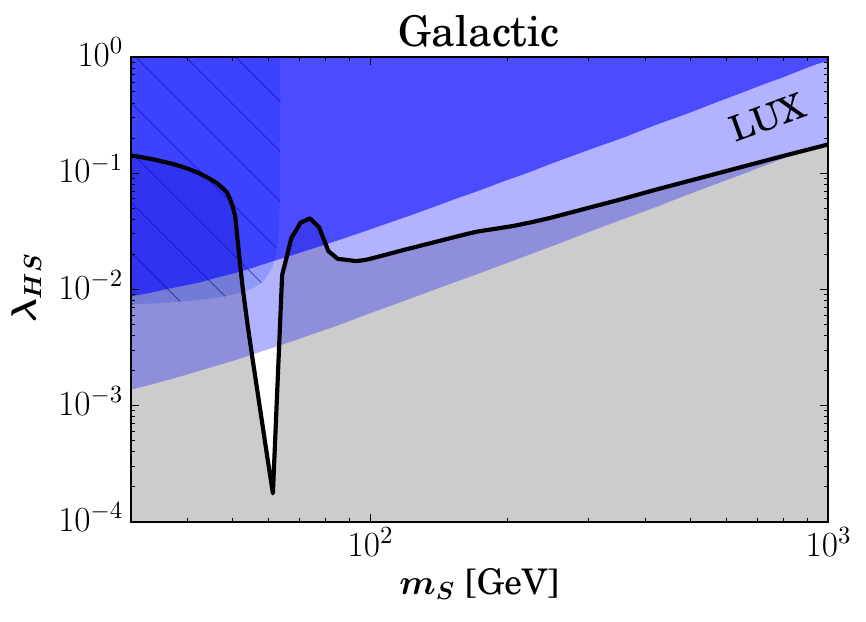}
\includegraphics[width=0.32\columnwidth]{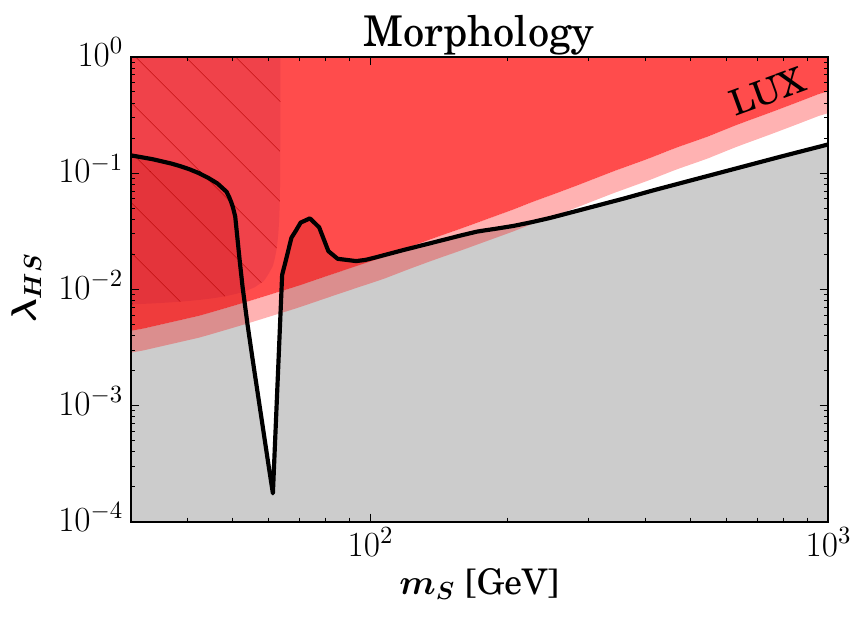}
\caption{\emph{Singlet Scalar Model}: Effects of Galactic uncertainties on the LUX exclusion limit in the SSDM parameter space.
We display the effect of: statistical uncertainty, for our reference morphology \refmorf~(left panel); varying the Galactic parameters for the same reference morphology (central panel); 
adopting different morphologies that maximize/minimize the local DM density $\rho_0$, {\it CjX} and {\it FiX}
(right panel).
Criteria are discussed in section~\ref{sec:astro} and values are reported in table~\ref{tab:barbrack}.
}
\label{fig:SSDM-1} 
\end{figure}

As it can be seen, the statistical uncertainty related to the determination of the local DM density $\rho_0$ affects the determination of model parameters very little, thus justifying the fact that most of the literature neglects it.
On the other hand, the uncertainty arising by either the ignorance about the exact value of Galactic parameters or the morphology of the visible component has sizable effects in shifting the constrained region in the parameter space. Notice that the largest uncertainty on the exclusion limit arises from the variation of the Galactic parameters. The reason is that this variation leads to a large variation in the value of $\rho_0$ (see the second and third rows of table~\ref{tab:barbrack}) which is larger than the variation in $\rho_0$ due to either statistical uncertainties or the choice of morphology, with the latter still being quite sizable, as we will discuss in the following.

\par The situation is different if one looks at the GeV excess favored region versus the constraint imposed by dSphs.
In figure~\ref{fig:SSDM-2}, we show the effect of Galactic uncertainties on the GeV excess favored region in the SSDM parameter space.
Blobs of different color shading are the regions that explain the GeV excess at 2$\sigma$ confidence level, shown as a red contour in figure~\ref{fig:GCvsdwarfs}, moving as a consequence of statistical, Galactic parameters, or baryonic morphology uncertainty. In this figure, for the baryonic morphology uncertainty, we choose the morphologies which maximise/minimise $\gamma$.
As it can be easily seen, again the statistical uncertainty on a single morphology plays little role, not affecting conclusions, but the adoption of different Galactic parameters and morphologies sizably shift the favored region, relieving (or worsening) tension with dSphs constraints, as it was already seen in figure~\ref{fig:GCvsdwarfs}.
It is interesting to notice that although the variation of Galactic parameters produces the most sizable alteration of the index $\gamma$ and of $\rho_0$ (for assigned morphology, see table \ref{tab:barbrack}), these effects are partially compensated in the computation of the \Jf, and the largest variation of the latter is obtained as a consequence of varying morphologies (for assigned Galactic parameters).

\begin{figure}[t!]
\centering
\includegraphics[width=0.32\columnwidth]{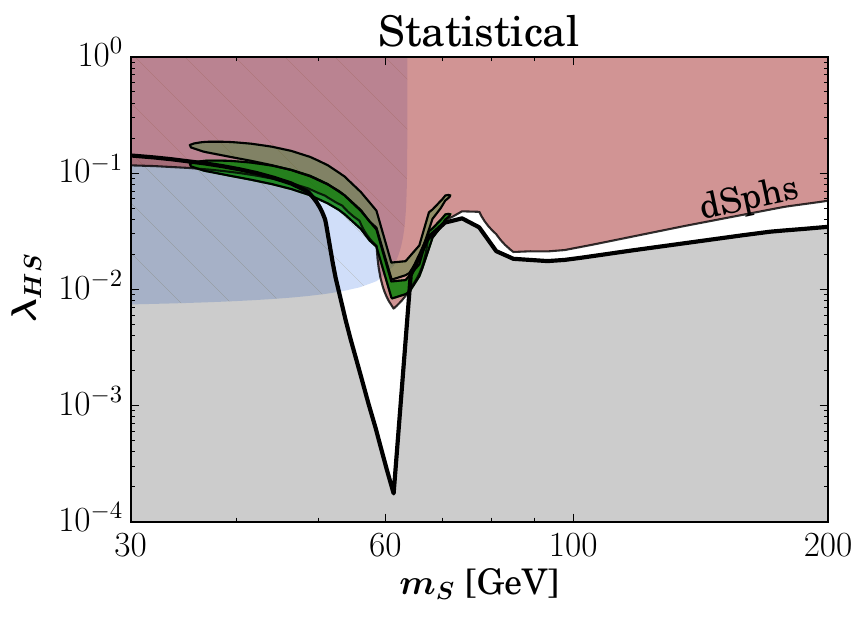}
\includegraphics[width=0.32\columnwidth]{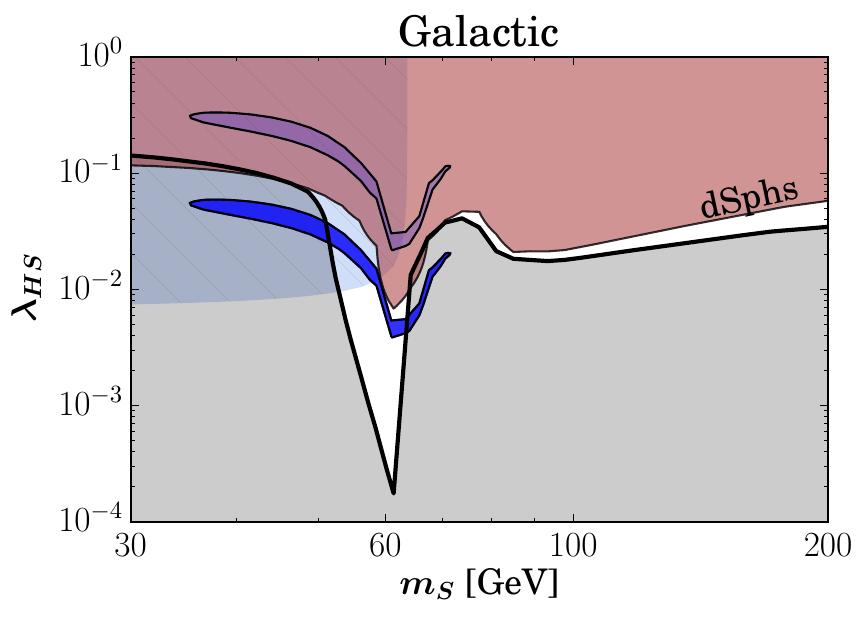}
\includegraphics[width=0.32\columnwidth]{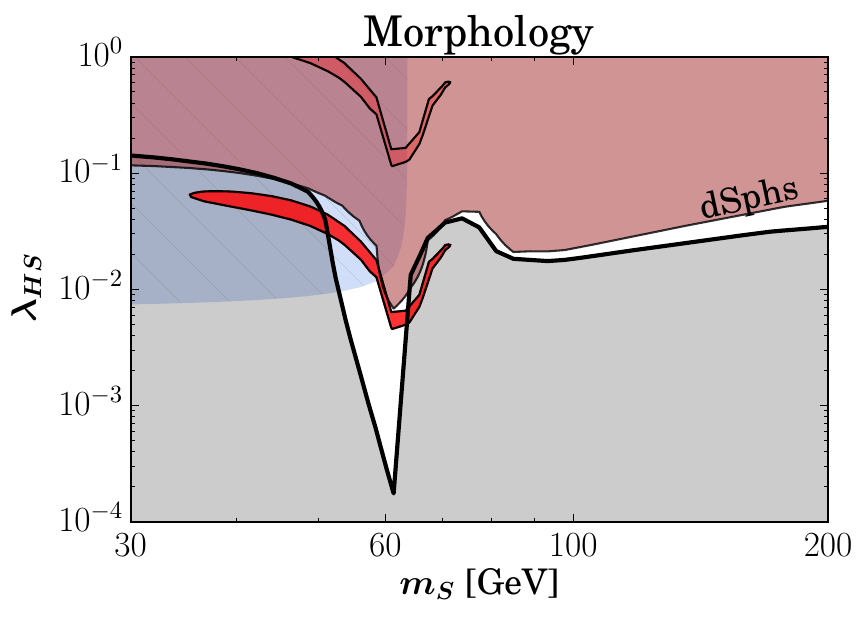}
\caption{\emph{Singlet Scalar Model}: Effects of Galactic uncertainties following the GeV excess interpretation.
We display the effect of: statistical uncertainty, for our reference morphology \refmorf (left panel); 
changing the Galactic parameters, for the same reference morphology (central panel); 
adopting different morphologies that maximize/minimize the index $\gamma$, {\it FkX} and {\it DiX}
(right panel).
Criteria are discussed in section~\ref{sec:astro} and values are reported in table~\ref{tab:barbrack}. 
}
\label{fig:SSDM-2} 
\end{figure}

\subsection{Impact on IDM Parameter Determination}

\par Following closely the procedure described in the previous section, now we compare the limit imposed by direct detection on the IDM parameter space for different cases of variation of astrophysical uncertainties: In figure~\ref{fig:IDM-1} we show how the LUX limit shown in the left panel of figure~\ref{fig:IDM} varies by including uncertainties arising from
{\it a)} the statistical uncertainty on our reference morphology, {\it b)} the variation of the Galactic parameters for the reference morphology, and {\it c)} the baryonic morphologies that maximize/minimize the local DM density $\rho_0$, as discussed in section~\ref{sec:astro}.
For DM direct detection, the effects of the systematic uncertainties on the Galactic parameters on the IDM are similar to the ones on the SSDM: On one hand, the statistical uncertainty related to the determination of the local DM density $\rho_0$ affects mildly the determination of model parameters.
On the other hand, the uncertainty arising by either the ignorance about the exact value of Galactic parameters or the morphology of the visible component has sizable effects in shifting the constrained region in the parameter space. As discussed before, the largest uncertainty in direct detection limits arises from the variation in the local DM density. The variation of the Galactic parameters for the reference morphology leads to the largest variation in $\rho_0$, and hence the largest uncertainty seen in the central panel of figure~\ref{fig:IDM-1}.
\begin{figure}[t!]
\centering
\includegraphics[width=0.32\columnwidth]{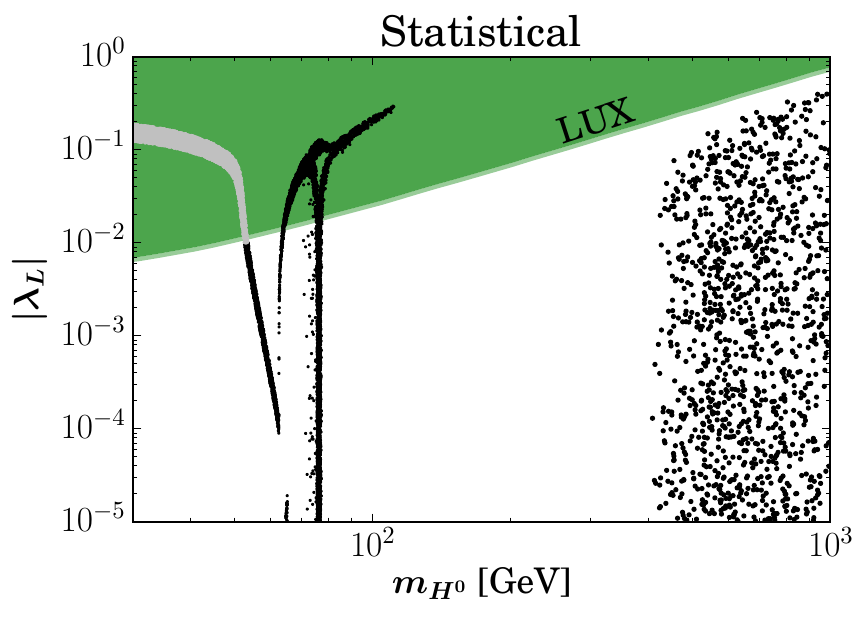}
\includegraphics[width=0.32\columnwidth]{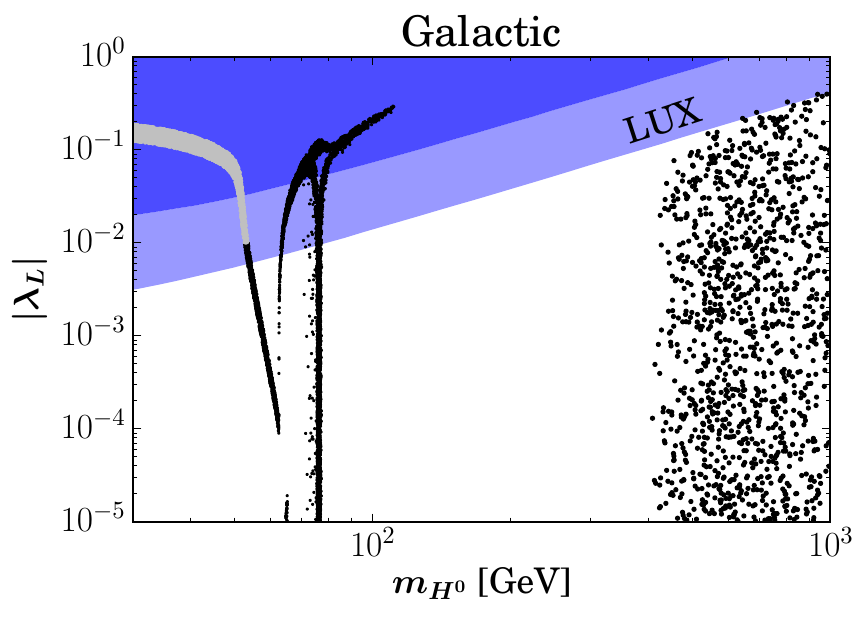}
\includegraphics[width=0.32\columnwidth]{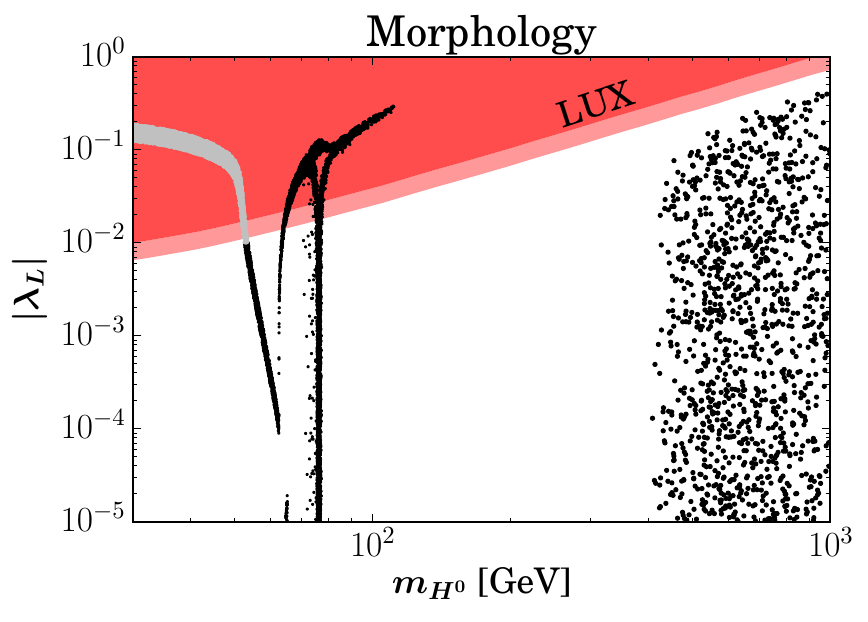}
\caption{\emph{Inert Doublet Model}: Effects of Galactic uncertainties on the parameter constraints.
We display the effect of: statistical uncertainty, for our reference morphology \refmorf (left panel); 
changing Galactic parameters, for the same reference morphology (central panel); 
adopting different morphologies such as they maximize/minimize the local DM density $\rho_0$, {\it CjX} and {\it FiX}
(right panel).
Criteria are discussed in section~\ref{sec:astro} and values are reported in table~\ref{tab:barbrack}. 
}
\label{fig:IDM-1} 
\end{figure}

\par In figure~\ref{fig:IDM-2}, we show how the GC excess interpretation shown in the right panel of figure~\ref{fig:IDM} varies with Galactic uncertainties. The dark gray dots show the constraint imposed by dSphs. 
Colored dots (green, blue and red) correspond to the regions of the parameter space that explain the GeV excess, moving as a consequence of statistical, Galactic, or morphology uncertainties, respectively. For the baryonic morphology uncertainty, we choose the morphologies which maximise/minimise $\gamma$.
It can be seen from the figure that the regions that can simultaneously reproduce the measured DM relic abundance and explain the GC excess are quite reduced and typically in tension with the dSphs observations.
Only marginal regions are allowed by the dSphs constraint when taking into account Galactic and morphological uncertainties.

It is to be noticed that figure~\ref{fig:IDM-2} displays only one region of the parameter space, as favored by the GC excess interpretation for both the Galactic parameters and the morphology, differently than in the case of SSDM.
The variation of both morphology and Galactic parameters impose a change in the IDM parameters (similar to what happens with the SSDM), but in both cases this shift ends up in a region which cannot reproduce the DM relic abundance. The ``shifted'' region is not visible in the figure as it is forbidden by the cosmological constraint, which henceforth practically sets limits on the DM interpretation of the GC excess, but only for some combinations of the Galactic parameters.

\begin{figure}[t!]
\centering
\includegraphics[width=0.32\columnwidth]{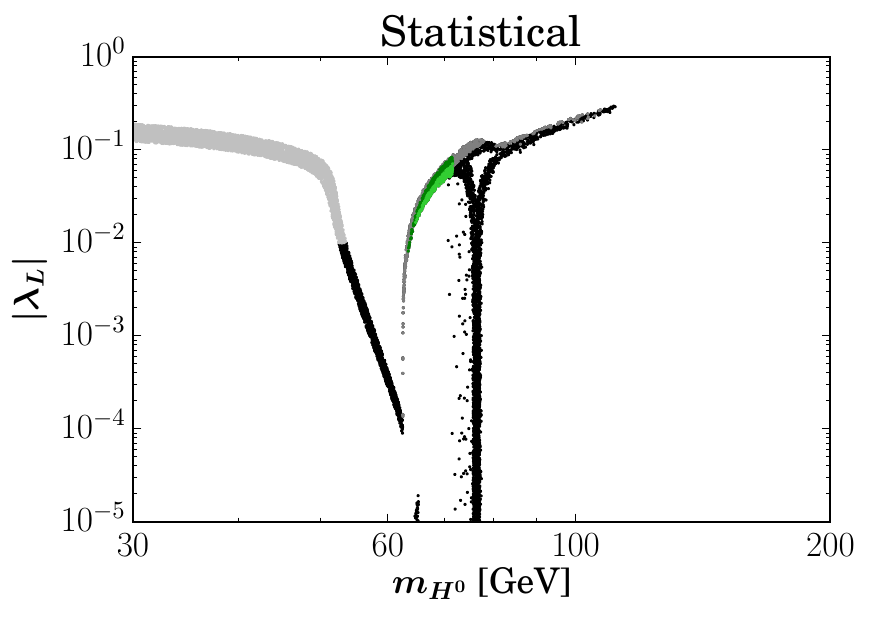}
\includegraphics[width=0.32\columnwidth]{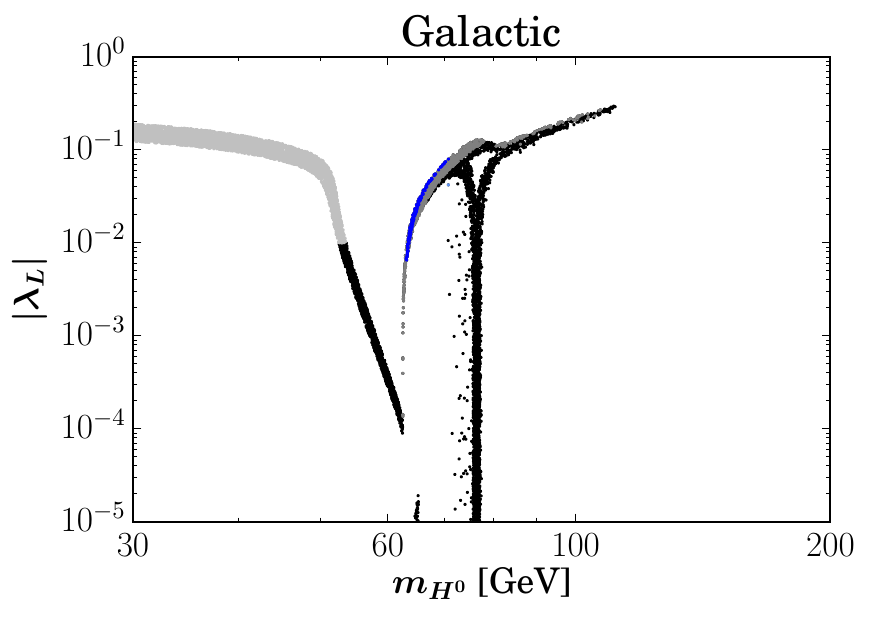}
\includegraphics[width=0.32\columnwidth]{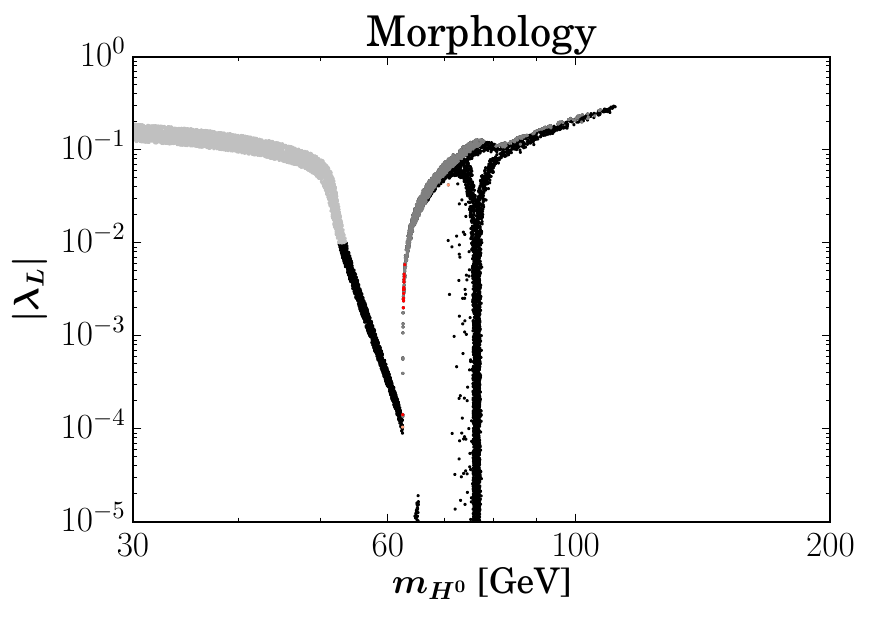}
\caption{\emph{Inert Doublet Model}: Effects of Galactic uncertainties following the GC excess interpretation.
We display the effect of: statistical uncertainty, for our reference morphology \refmorf~ (left panel); 
changing Galactic parameters, for the same reference morphology (central panel); 
adopting different morphologies such as they maximize/minimize the index $\gamma$, {\it FkX} and {\it DiX}
(right panel).
Criteria are discussed in section~\ref{sec:astro} and values are reported in table~\ref{tab:barbrack}. 
The colored dots (green, blue and red) correspond to the regions of the parameter space that explain the GC GeV excess; the dark gray dots are in tension with the constraint imposed by dSphs. 
}
\label{fig:IDM-2} 
\end{figure}

\section{Conclusions}\label{sec:conc}

\par In this work we have studied how the uncertainties associated
to Galactic core quantities, such as the local
galactocentric distance, local circular velocity, and the morphology of the stellar disk and bulge, affect the determination
of DM distribution, and eventually propagate when constraining new physics scenarios.
We have set up a systematic scan of the major sources of uncertainty in the determination of 
the DM distribution in the MW, testing (a) statistical uncertainties; (b) variation of Galactic parameters; (c) variation of baryonic morphology.
While the purely statistical uncertainties affecting the observed RC
and the normalization of the visible mass component do not sizably affect the
constraints on new physics model parameters, 
a significant impact on the allowed model parameter space is due to the 
current ignorance on the morphology of the baryonic component, and on the
determination of Galactic parameters.
\par We have shown that the latter significantly affect the 
constraints of two specific models, the SSDM and the IDM, which we have 
chosen as testbeds for the relatively simple dependence of their phenomenology on the
key parameters. Our main findings, which we summarize below, show the need for 
the study of these uncertainties in more complex scenarios, 
and an increased communication between the particle physics and the astronomy communities in a virtuous interplay.
\par The largest effects on the SSDM and IDM parameter space are obtained as a consequence of varying the Galactic parameters ($R_0$, $v_0$):
The variation of ($R_0$, $v_0$) between its currently established extreme values pushes the determination 
of the local DM density $\rho_0$ beyond the usually
adopted bounds (which are taken for assigned Galactic parameters, and include statistical uncertainty only, in most
 cases), with major effects especially on direct detection results. Interestingly, the remarkable changes imposed
by the Galactic parameter variation also on the index $\gamma$ mitigate the effect on the determination of the \Jf,
which sees the uncertainty on the baryonic morphology as a primary source of uncertainty for indirect detection.
\par As an example of the above, we recall here the case of SSDM: The region of the parameter space
which permits an interpretation of the GC excess in terms of DM annihilation is allowed with a given set
of Galactic parameters, but it could be also entirely ruled out by constraints on the relic density if the other extreme values for ($R_0$, $v_0$)
are adopted.
\par Accounting for the astrophysical uncertainties described above, will be even more crucial in the case a 
tantalising DM signal will be discovered in the next--generation of direct and indirect experiments. In that case, the accurate reconstruction and interpretation of 
the signal in the context of concrete particle physics models will require the full treatment of all astrophysical uncertainties presented in our work.

\par On the other hand, future astronomical data will help in reducing significantly those uncertainties. 
In particular, the \emph{Gaia} mission is expected to improve the determination of the Oort constants, and yield a reduction of uncertainties on the determination of ($R_0$, $v_0$), 
as already shown possible with the first year data release~\cite{2016arXiv161007610B}.


\vspace{0.5cm}
{\it Acknowledgments.}
We thank P.~D.~Serpico for fruitful discussion and comments on the manuscript.
N.~Bernal is supported by the S\~ao Paulo Research Foundation (FAPESP) under grants 2011/11973-4 and 2013/01792-8, by the Spanish MINECO under Grant FPA2014-54459-P and by the `Joint Excellence in Science and Humanities' (JESH) program of the Austrian Academy of Sciences. 
N.~Bozorgnia acknowledges support from the European Research Council through the ERC starting grant WIMPs Kairos. 
F.I.~is supported by FAPESP JP project 2014/11070-2. 
This research has been made possible through the FAPESP/GRAPPA SPRINT agreement 2015/50073-0. 
The authors would like to express a special thanks to the Mainz Institute for Theoretical Physics (MITP) for its hospitality and support. 


\bibliographystyle{JHEP}
\bibliography{astrounc}

\end{document}